\documentclass[aps,preprint,superscriptaddress]{revtex4}
\usepackage{psfrag,graphicx}

\newcommand{\Bra}[1]{\left \langle #1 \right |} 
\newcommand{\Ket}[1]{\left | #1 \right \rangle}  
\newcommand{\BraKet}[2]{ \langle #1 | #2  \rangle} 
\newcommand{\ME}[3]{\Bra{#1} #2 \Ket{#3}} 
\newcommand{\RME}[3] {\langle #1 || #2 || #3  \rangle}  
\newcommand{\sixj}[6]{\left \{ \begin{array}{ccc} #1 & #2 & #3\\
                        #4 & #5 & #6 \end{array} \right \} }
\newcommand{\abs}[1]{\left \arrowvert #1 \right \arrowvert }
\newcommand{\ninj}[9]{\left \{  \begin{array}{ccc} #1 & #2 & #3\\
                            #4 & #5 & #6\\  #7 & #8 & #9
                      \end{array} \right \}      }
\begin{document}
         
\title{A simple model for $f\rightarrow d$ transition of heavy lanthanide and
 actinide ions in crystals}

\author{Chang-Kui Duan}
\affiliation
{Institute of Modern Physics,
Chongqing University of Post and Telecommunications, Chongqing 400065,
China}
\affiliation{Department of Physics and Astronomy, University of
  Canterbury,  Christchurch,  New Zealand}
\author{Michael F. Reid}
\affiliation{Department of Physics and Astronomy, \\and MacDiarmid
  Institute  of Advanced Materials and Nanotechnology,\\ University of
  Canterbury, Christchurch,  New Zealand}
\author{Gang Ruan}
\affiliation
{Institute of Modern Physics,
Chongqing University of Post and Telecommunications, Chongqing 400065,
China}

\begin{abstract}
The $f\rightarrow d$ transition model by Duan and co-workers 
[Phys. Rev. B {\bf 66}, 155108 (2002); J. Solid State Chem. {\bf 171}, 299 (2003)]
has been very useful in interpreting the $f\rightarrow d$ absorption,
emission and nonradiative relaxation of light lanthanide ions in crystals.
However, based on the assumption that the $f^{N-1}$ core spin-orbit 
interaction is weak compared to $f\rightarrow d$ exchange interaction,
this model, in the original form, is not applicable to interpretation of
 the $f\rightarrow d$ transitions of  heavy 
lanthanide ions or actinide ions in crystals. In this work the model is extended 
to cover the cases of heavy lanthanides and actinides, where the spin-orbit
interaction of $f$ orbitals may be stronger than the $f-d$ exchange interaction.

\end{abstract}

\maketitle

\section{Introduction}

 New luminescent phosphors for vacuum ultraviolet (VUV)
 excitation are required for plasma display panels 
and mercury-free fluorescent tubes, where the VUV emission
from a noble gas xenon discharge is used to generate
visible luminescence. Other applications where the
VUV spectroscopy of lanthanides is involved are
scintillator materials and VUV lasers. Due to these potential
applications and availability of VUV excitation by
synchrotron radiation, the VUV spectroscopy of lanthanide ions
and actinide ions in crystal have recently become an
important field of research. 

The VUV spectroscopy of lanthanide and actinide  ions mainly involves
the parity allowed $nf^N\leftrightarrow nf^{N-1}(n+1)d$ transitions
($n=4$ for lanthanides and 5 for actinides), including ground 
and excited state excitations and emissions.
It is well known  that the $4f^ N \rightarrow 4f^ N$
 spectra of lanthanide ions  in noncentrosymmetric environments 
are dominated by   many sharp zero-phonon lines.
 Interpretation of the energy
levels and transition intensities may be modeled with a empirical crystal-field
 Hamiltonian \cite{New2000} and Judd-Ofelt
 theory\cite{Jud1962,Ofe1962} respectively.
 By contrast, the spectra of transitions between the
 $4f^N$ and $4f^{ N-1}5d$ configurations are feathered by
 broad-band structures, with some zero-phonon lines
only resolvable at temperature below liquid nitrogen. This 
is due to great difference in vibrational equilibrium 
between $4f^{N-1}5d$ and $4f^N$ configurations,
which makes the the transition intensities be
dominant by vibronic broad bands. The phenomenological 
crystal-field Hamiltonian for $4f^N$ configuration
has been extended to $4f^{N-1}5d$ configuration by
 adding the crystal-field and spin-orbit interactions
for $5d$ orbitals, and Coulomb interactions between
 $4f^{N-1}$ core and $5d$ orbitals\cite{Rei2000}. The 
$4f^N\leftrightarrow 4f^{N-1}5d$ transitions 
are electric dipole allowed, whose relative rates
 can be calculated straightforward. Actually, extensive
 calculations have been carried out for trivalent
 lanthanide ions in crystals \cite{Rei2000,Tan2003}
which give satisfactory agreement with experimental
 $4f \rightarrow  5d$ spectra. However, the calculations give 
hundreds to thousands transition lines which convolve
into several to a few dozen of broad bands after taking
vibronic transitions into account to wash out the 
fine structures, making straightforward interpretations of
the simulations and straightforward
predictions of changing of broad-band numbers, their 
positions and relative intensities with ions and crystals
quite difficult.

 Recently Duan and co-workers \cite{DuaR2002,Dua2003,Lin2004} simplified the 
calculations by considering only the main interactions
in the  $4f ^{N-1}5d$  configuration. The model gave quantum
numbers characterizing various groups of states,
 transition selection rules, and an expression of 
relative intensities with these quantum numbers. Application of
 the model to Eu$^{2+}$ and Sm$^{3+}$ in various crystals successfully 
explained the measured spectra. However, in the model there is
an implicit assumption that the exchange Coulomb interaction
between $f$ and $d$ orbitals is stronger than the spin-orbit
interaction in the $f^{N-1}$ core. This assumption no long holds
for heavy lanthanide ions or actinide ions in crystals. 

In the present work, the original model is
extended to the  case where $f-d$ exchange
Coulomb interaction may be weaker 
than that of the $f^{N-1}$ core spin-orbit interaction.
The energy and transition line strengths  
for $f-d$ transitions directly applicable to actinides and
heavy lanthanides  are given analytically.

\section{Eigenfunctions and eigenvalues}

\subsection{${\lowercase{f}}^N$ configuration}

The $f^N$ crystal-field splitting is well-understood both experimentally
 and theoretically via phenomenological 
 crystal-field simulation. However, in $f^N\rightarrow f^{N-1}d$ transitions, 
the  $f^N$ crystal-field energy levels are either hard to distinguish from
 vibronic lines in low temperature or unresolvable at all at temperature
 higher than 100K due to vibronic bands. In this work we are aimed to give an 
model which may  interpret  number of broad bands, their positions and relative 
 intensities. We neglect $f^N$ Crystal-field splitting.
 In such a case, the highly degenerate eigenstates can be
 written as
 \begin{equation}
 \Ket{[\eta S L ] J} = C_0 \Ket{\eta S L J} 
                     + \sum\limits_{i\geq 1} c_i \Ket {\eta_i S_i L_i J }.                   
\end{equation}
Usually the states are denoted with the label of the main component
$\Ket{\eta S L J}$, whose coefficient is usually close to 1 for lanthanide ions.

\subsection{$\lowercase{f}^{N-1}\lowercase{d}$ configuration}

The general interactions for the $f^{N-1}d$ configuration can be written as
\begin{equation}
\label{fullH}
H(f^{N-1}d) = H_{\rm Coul}(ff) + H_{\rm cf}(d) + H_{\rm Coul} (fd) + H_{\rm 
so}(f)            + H_{\rm so}(d) + H_{\rm cf}(f).
\end{equation}
The first two terms are the strongest terms of the following
form
\begin{equation}
H_0(f^{N-1}d) = \sum\limits_{k=2,4,6} F_k(ff) \sum\limits_{1\leq i<j \leq N-1} 
C^k(i)\cdot C^k(j)+\sum\limits_{k=2,4}\sum\limits_{-k\leq q \leq k} B^k_q 
C^{(k)}_q(d),
\end{equation}
where $F_k(ff)$ are slater integrals, which are usually treated as adjustable 
parameters. $B^k_q$ are crystal-field parameters for the $d$ electron. Note that only
 those  $B^k_q$ with $(k,q)$ allowed by the site symmetry are nonzero.
$H_0(f^{N-1}d)$ contribute to splitting of $f^{N-1}$ core into energy 
levels  characterized with spin and orbit 
angular momenta $S$ and $L$, and $d$ orbital into strong crystal-field energy levels
 characterized with  site-symmetry labels.
The  contribution to splitting from the third term of (\ref{fullH}),
 {\it i.e.}, the Coulomb interaction between $f^{N-1}$ core and $d$ orbitals,
 can be approximate with an isotropic exchange term
\begin{eqnarray}
&&H_{\rm exc}(f^{N-1}d) = -J_{\rm exc} {\bf S_f} \cdot {\bf S_d},{~~\rm 
where}\nonumber \\
&&J_{\rm exc} = \frac{6}{35} G_1(fd) +\frac{8}{105} G_3(fd)+ 
\frac{20}{231} G_5(fd).
\end{eqnarray}
Here $G_{1,3,5}(fd)$ are $f-d$ Coulomb exchange radial integrals. 
It is straightforward to show
that $H_{\rm exc}$ commutes with the total spin of $f^{N-1}d$.

The following approximation is often used for $f$-electron spin-orbit 
interaction within a given zero-order $f^{N-1}$ core energy level characterized by 
$\eta S L$:
\begin{eqnarray}
\label{hso}
H_{\rm so}(f) &=& \sum\limits_{i=1}^{N-1}\xi_{nl} {\bf s_i}\cdot {\bf 
l_i}\approx \lambda_{\eta S L}{\bf S_i} \cdot {\bf L_i}, {~~\rm 
where}\\ 
\lambda_{\eta S L } &=& \sqrt {\frac{l(l+1)(2l+1)}{S(S+1)(2S+1)L(L+1)(2L+1)}}
                     \RME{l^N \alpha SL}{V^{11}}{l^N \alpha SL} \xi_{nl}.
\end{eqnarray}
In the case that $S_f$ takes the largest possible value for the $f^{N-1}$ 
configuration,  $\lambda_{\eta_f S_f L_f}$ is simply  ${\rm sign}(8-N)\xi_{f}/2S_f$
\cite{Wyb1965,Dua2003}. In general, $H_{\rm so}(f)$ commutes with the total angular 
momentum of $f^{N-1}$ core, no matter the above approximation in Eq. (\ref{hso}) is used
or not.

Other terms are not important in the interpretation of 
broad bands in $f-d$ spectra and neglected. Therefore we have the effective Hamiltonian 
written as the sum of the above important terms as
\begin{equation}
H_{\rm eff} = H_0(f^{N-1}d) - H_{\rm exc} (f^{N-1}d) + H_{\rm so}(f).
\end{equation}
It is straightforward to check that $H_{\rm eff}$ commutes with the following 
effective ``angular momentum'' operator
\begin{equation}
{\bf J_{\rm eff}} = {\bf S}_f + {\bf S}_d + {\bf L}_f.
\end{equation}
Note that this operator is not the total angular momentum operator for 
$f^{n-1}d$, since it does not contain the orbit angular momentum of $d$ 
electron which is usually quenched in low symmetry sites.
 
 Former work used an implicit assumption that $fd$ exchange interaction is 
stronger than  $f^{N-1}$ spin-orbit interaction. In that case the energy levels 
were written as
\begin{eqnarray}
&& E \left (\Ket{l^{N-1} \eta S_f L_f,\ ^2 d_i; S J}\right )
= E_0(\eta S_f L_f) + \epsilon_{d_i} 
- J_{\rm exc}\left [S(S+1-S_f(S_f+1)-\frac{3}{4}\right ] 
\nonumber\\
&&
+  \left (2-\frac{2S+1}{2S_f+1}\right ) \lambda_{\eta_f S_fL_f} 
    \frac{J(J+1)-S(S+1)-L_f(L_f+1)}{2}.
\end{eqnarray}
However, the strength of the exchange interaction decreases as the nucleus charge 
increase, while at the same time the spin-orbit interaction increases. It happens 
that for heavy lanthanide and actinide ions, the cases where spin-orbit 
interaction is comparable or even stronger than exchange interaction need to be 
considered.
In the case $H_{\rm so}(f)$ is much stronger than $H_{\rm exc}(fd)$, 
 opposite to the one considered by Duan {\it et. al}\cite{Dua2003}, The coupling 
$(SO_3^{S_f}\times SO_3^{L}) \times SO_e^{S_d}$ may be preferred and the approximate 
eigenstates can be written as $\Ket{(l^{N-1}\eta S_f L_f, J_f), ~^{2}d_i; J }$ 
and the eigenvalues can be written as 
\begin{eqnarray}
&& E(\eta S_f L_f J_f, ~^2d_i;J) = E_0 (\eta S_f L_f, ~^2d_i)
 +\lambda _{\eta S_f L_f}\left [ \frac{(J_f(J_f+1)-S_f(S_f+1)-L_f(L_f+1)}{2}\right ]
 \nonumber\\
&&~~~- J_{\rm exc} \frac{J_f(J_f+1)+S_f(S_f+1)-L_f(L_f+1)}{2J_f(J_f+1)}
\left [ J(J+1)-J_f(J_f+1)-S_d(S_d+1)\right ].
\end{eqnarray}
In the medium case where $H_{\rm so}(f)$ and $H_{\rm exc}(d)$ are comparable, 
The effective of $H_{\rm eff}$ can be calculated with either 
$\Ket{l^{N-1}\eta S_f L_f J_f, ~^2d_iJ}$ or $\Ket{l^{N-1}\eta S_f L_f, ~^2d_i S 
J}$as bases. Eigenvalue and wave-function of each eigenstate can then be
obtained by diagonalizing the matrix.  Here we give the matrix element of 
effective Hamiltonian under the bases $\Ket{l^{N-1} \eta S_f L_f ~^2d_i SJ}$ as
\begin{eqnarray}
&&\ME{l^{N-1}\eta S_f L_f, ~^2d_i; SJ}{H_{\rm eff}}{ l^{N-1}\eta S_f L_f, 
~^2d_i; S'J'}\nonumber\\
&=&\delta_{d_i,d_i^{\prime}} 
\{\delta_{S_fS_f^{\prime}}\delta_{L_fL_f^{\prime}}\delta_{SS'}[(E_0(f^{N-1}\eta 
S_f L_f) + \epsilon_{d_i})
-J_{\rm exc} \frac{S(S+1)-S_f(S_f+1)-s_d(s_d+1)}{2}] \nonumber\\
&+&\xi_{nl}(-1)^{J+L_f+S_f^{\prime}} 
     \sixj {L_f} {L_f^{\prime}} 1 {S^{\prime}} S J 
     \sqrt{l(l+1)(2l+1)} \RME {\eta S_fL_fSJ}{V^{11}}
     {\eta^{\prime}S_f^{\prime}L_f^{\prime}S^{\prime}J}
\label{medium}
\end{eqnarray}
Under the approximation in Eq.(\ref{hso}), the matrix for $H_{\rm eff}$ 
reduces into many $2\times 2$ blocks and the diagonization become straightforward.

\section{One-photon transition line strength between $\lowercase{f}^N$ and
 $\lowercase{f}^{N-1}\lowercase{d}$}

The $f^N$ to $f^{N-1}d$ transitions are electric dipole allowed. Here we consider
only this mechanism.  The electric dipole moment
 is a spin independent rank 1 tensor in both total orbital
 angular momentum and total angular momentum spaces. It can be
 written as
\begin{eqnarray}
D = \sum\limits_q \epsilon_q \sum\limits_{i=1}^N r_q(i),
\end{eqnarray}
where $\epsilon_q$ is the $q$ component of the polarization
 vector and $r_q(i)$ is the $q$ component of the position
of $i^{\rm th}$ electron. 

Using the second quantization techniques, the electric dipole momentum 
can be written as
\begin{eqnarray}
D^{(0.1)1}_q =  \sqrt{2}\ME{f}{r}{d} 
  \left \{ [(a^+)^{(\frac{1}{2}\cdot 3)}\tilde{a}^{(\frac{1}{2}\cdot 2)}]^{(0\cdot1)1q}- 
  [(a^+)^{(1/2\cdot2)}\tilde{a}^{(1/2\cdot 3)}]^{(0\cdot 1)1q}\right \},
\label{ed}
\end{eqnarray}
where 
\begin{eqnarray}
(\tilde{a})^{sm_slm_l} = (-1)^{s-m_s+l-m_l} a^{sm_slm_l},
\end{eqnarray}
and $(a^+)^{(sm_slm_l)}$ are components of tensors that transform under symmetry operator
the same way as basis $\Ket{sm_slm_l}$, and $\langle f |r|d \rangle$ is radial integral.
The coupling of two
creation-annihilation operators are just coupling of two tensors  to give
 a new tensor. In Eq.(\ref{ed}), $a^+$ and $\tilde{a}$ couples to give a rank $(0\cdot 1)$
tensor of spin and orbital angular moments.

Using coupling and recoupling techniques, we can rewrite  the electric dipole
momentum  into the following two forms.
\begin{eqnarray}
\label{form1}
D    &=& \sum\limits_{q_1,q_2} C^1_{q_1q_2}
[(a^+)^{(1/2\cdot 3)}\tilde{a}^{(1/2\cdot 2)}]^{(0\cdot 3 \cdot 2)(3 q_1,2q_2)}
+\cdots
\\
&=& \sum\limits_{j_f,q_1,q_2} C^2_{j_fq_1q_2} 
[(a^+)^{(1/2\cdot 3)}\tilde{a}^{(1/2\cdot 2)}]^{(j_f\cdot 1/2\cdot 2)(3q_1,2q_2)}
+\cdots,
\label{form2}
\end{eqnarray}
where the neglected terms ($\cdots$) will not contribute
 when $f^N$ states are on the left and
$f^{N-1}d$ are on the right and hence are not written ou
t explicitly. $C^1_{q_1q_2}$ and $C^2_{j_fq_1q_2}$
are appropriate coefficients that depend on $(q_1,~q_2)$
 and $(j_f,~q_1,~q_2)$ respectively.
The matrix elements of $D$ between initial states
$\Ket{f^Nd^0 SLJ}$ and final states
 $\Ket{(f^{N-1}d^1 ((S_fs_d)SL_f)J^{\prime},d_i}$
( or final states $\Ket{(f^{N-1}d^1((S_fL_f)J_fs_d)J^{\prime},d_i}$)
 can then be obtained
{\it via} Wigner-Ekwart theorem\cite{But1981}.

Transition line strength $S$ between initial states $\Ket{Ii}$ and
$\Ket{Ff}$, where $i$ and $f$ are indexes to distinguish 
degenerate states, are defined as follows
\begin{eqnarray}
S(I,F) = \sum\limits_{i,f}\abs{\ME{Ii}{D}{Ff}}^2.
\end{eqnarray}
Using orthonormal relations of coupling and recoupling coefficients\cite{But1981}, after
 a lengthy but straightforward 
analytic calculation,  we finally get the line strength for isotropic
absorption or emission. For the case I of strong $H_{\rm exc}(f^{N-1}d)$ which is
applicable to light lanthanides, the line strength is:
\begin{eqnarray}
&&S_{\rm iso} \left ( \Ket{f^N\eta SLJ}\leftrightarrow 
               \Ket{[(f^{N-1}\eta _fS_fL_f,~^2d_i; S^{\prime}J^{\prime}}\right )
\nonumber\\
&&~~~~= \frac{N\ME{f}{r}{d}^2}{35} 
 \delta_{SS^{\prime}} [d_i][L,J,J^{\prime}]
 \BraKet{f^N\eta SL\{}{f^{N-1}\eta_fS_fL_f}^2.
    \sixj{L}{L_f}{3}{J^{\prime}}{J}{S^{\prime}}^2.
\end{eqnarray}
where the $[d_i]$ is the degeneracy of $d_i$ crystal field levels, and
$[S]$ {\it etc.} are short for $(2S+1)$ {\it etc.}, respectively.

For the case II of stronger $H_{\rm so}(f)$ than $H_{\rm exc}(f^{N-1}d)$,
the line strength is:
\begin{eqnarray}
&&S_{\rm iso}(\Ket{f^N\eta SLJ}\rightarrow
                  \Ket{[(f^{N-1}\eta_f S_fL_f)J_f,~^2d_i;J^{\prime}})
= N\ME{f}{r}{d}^2[d_i][S,L,J_f,J,J^{\prime}]\nonumber
\\
&&\times 
    \BraKet{f^N\eta SL\{ }{f^{N-1}\eta_fS_fL_f}^2
    \left (\sum\limits_{j_f=5/2}^{7/2}(-1)^{j_f-7/2}[j_f]
    \ninj{1/2}{S_f}{S}{3}{L_f}{L}{j_f}{J_f}{J}
    \sixj{j_f}{3}{1/2}{J^{\prime}}{J_f}{J}
    \right )^2,
\end{eqnarray}

\section{Conclusion}

The model for $f^N\rightarrow f^{N-1}d$ transitions proposed
earlier\cite{Dua2003}  has been extended to the case II where spin-orbit
interaction in $f^{N-1}$ is stronger than
the isotropic exchange interaction between $f^{N-1}$ and $d$ by utilizing
 Racah-Wigner algebra and second quantization techniques. The result is expected
to be useful for actinide ions where the effect of $f$ spin-orbit interaction is
 stronger than the exchange interaction.
Heavy lanthanides fall into the medium case where the effect of spin-orbit interaction
is compariable or slightly stronger than the exchange interaction, where the case 
II may serve as an approximaition.

\newpage

\bibliography{fd3}

\end{document}